\documentclass[A4paper,preprint,12pt,tightenlines,amsmath,amssymb,floatfix,nofootinbib,aps,raggedbottom]{revtex4-2}

\usepackage{graphicx}
\usepackage{bm}
\usepackage{url,multirow,xcolor}
\usepackage[font=footnotesize,margin=24pt,justification=centerlast]{caption}
\usepackage{xtab}

\begin{document}

\title{ Vertex correction to nuclear matrix elements of double-$\bm{\beta}$ decays \vspace{10pt} }

\author{J.\ Terasaki \\ \vspace{0pt}}
\affiliation{ Institute of Experimental and Applied Physics\hbox{,} Czech Technical University in Prague, Husova 240/5, 110\hspace{3pt}00 Prague 1, Czech Republic \vspace{10pt}}


\begin{abstract} 
\vspace{3pt}
The predicted neutrinoless double-$\beta$ ($0\nu\beta\beta$) decay is the crucial phenomenon to prove the existence of the Majorana neutrino, which gives a foundation to leptogenesis to explain the matter prevalence of the universe. The nuclear matrix element (NME) of $0\nu\beta\beta$ decay is an important theoretical quantity to determine the effective neutrino mass and help the detector design for the next generation of the $0\nu\beta\beta$ decay search. Reliable calculation of this NME is a long-standing problem because of the diversity of the predicted values of the NME. The main reason for this difficulty is that the effective strength of the Gamow-Teller transition operator $g_A$ for this decay is unknown. I will show the lowest-order vertex corrections for the $0\nu\beta\beta$ and the $2\nu\beta\beta$ NME of $^{136}$Xe in the framework of the hybrid application of the quantum field theory to the leptons and the Rayleigh-Schr\"{o}dinger perturbation to the nucleus. The unperturbed nuclear states are obtained by the quasiparticle random-phase approximation. These corrections reduce the $0\nu\beta\beta$ NME by 30\%. The effective $g_A$ referring to this reduced NME is also obtained, and it is shown for the first time that the effective $g_A$ for the $0\nu\beta\beta$ NME is not quite different from that for the $2\nu\beta\beta$ NME; the difference is only 10\%. This indicates the possibility that the phenomenological effective $g_A$ to reproduce the experimental half-life of the $2\nu\beta\beta$ decay can be approximately used for the calculation of the $0\nu\beta\beta$ NME. 
\end{abstract}
 
%
\maketitle
\newpage
\section{\label{sec:introduction}Introduction}
Majorana neutrino is one of the hypothetical particles playing an important role in leptogenesis to explain the matter prevalence of the universe \cite{Fuk86}. The Majorana neutrino search is a major subject in modern physics because of this importance. A Majorana neutrino is a self-conjugate particle. This exotic property enables the exchange of the Majorana neutrino between nucleons, and this interaction causes the neutrinoless double-$\beta$ ($0\nu\beta\beta$) decay \cite{Fur39}; this decay is possible if and only if the Majorana neutrino exists. Many experimental groups conduct searches for this decay to find evidence of the existence of the Majorana neutrino \cite{Mer24}. The signal of this decay is two emitted electrons with the energy at the upper endpoint of the two-electron spectra of the weak decay with the usual -neutrino emission. The $0\nu\beta\beta$ decay also proves, if found, the lepton number nonconservation. The decay rate is proportional to squared the effective neutrino mass (Majorana mass) $\langle m_\nu \rangle$, which is unknown for now due to the new parameters called the Majorana phase in the extension of the Pontecorvo-Maki-Nakagawa-Sakata (PMNS) matrix \cite{Pon57,Mak62}. If the half-life of the $0\nu\beta\beta$ decay is measured and the reaction matrix element is given, $\langle m_\nu \rangle$ can be determined. This is another goal of the Majorana neutrino search. The importance of this goal is because the neutrino mass scale is not quantitatively known and this is one of the historical problems of the neutrino. 

A feature of the $0\nu\beta\beta$ decay is that nuclei are necessary because of the necessity of the positive Q value. Thus, nuclear physics is necessary to calculate the reaction matrix element. It is known that the factor depending on the nuclear wave functions, called the nuclear matrix element (NME), strongly depends on the calculation methods. The maximum-to-minimum ratio of the calculated values is 2.5$-$5 \cite{Ago23} for several nuclei with $A$ $\ge$ 76. The accurate NME is necessary for the determination of $\langle m_\nu \rangle$ and the design of the new detectors. When reasonable nuclear wave functions are used for the calculation of the Gamow-Teller (GT) transition strength to reproduce the measured value, the effective value of the strength of the GT transition operator $g_A$ is necessary, which is appreciably smaller than the bare value of 1.27 \cite{Bro85}. It is unknown whether that effective value can be used for the $0\nu\beta\beta$ decay. This is the major cause of the problem with the NME. The difference between the effective and bare values indicates that a correction to the transition operator is necessary due to the many-body correlations. In this article, I calculate the lowest-order correction, and the perturbed effective $g_A$ is obtained, which reproduces the NME with the perturbed transition operator and the bare $g_A$. Here, the unperturbed transition operator is used for the calculation with the perturbed effective $g_A$. The equations are shown in Sec.~2, and the results of the calculation are shown in Sec.~3. The implications of the result are also discussed.

\section{\label{sec:eq_diag} Equations and Diagrams}
\begin{figure}[t]
\includegraphics[height=0.2\columnwidth]{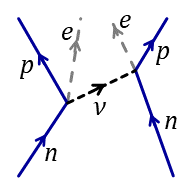}
\vspace{-10pt}
\caption{ \protect \label{fig:0vbb_0th} \baselineskip=13pt 
Diagram of unperturbed $0\nu\beta\beta$ decay. Those parts used in the diagram are proton ($p$), neutron ($n$), electron ($e$), and Majorana neutrino ($\nu$).}

\includegraphics[height=0.25\columnwidth]{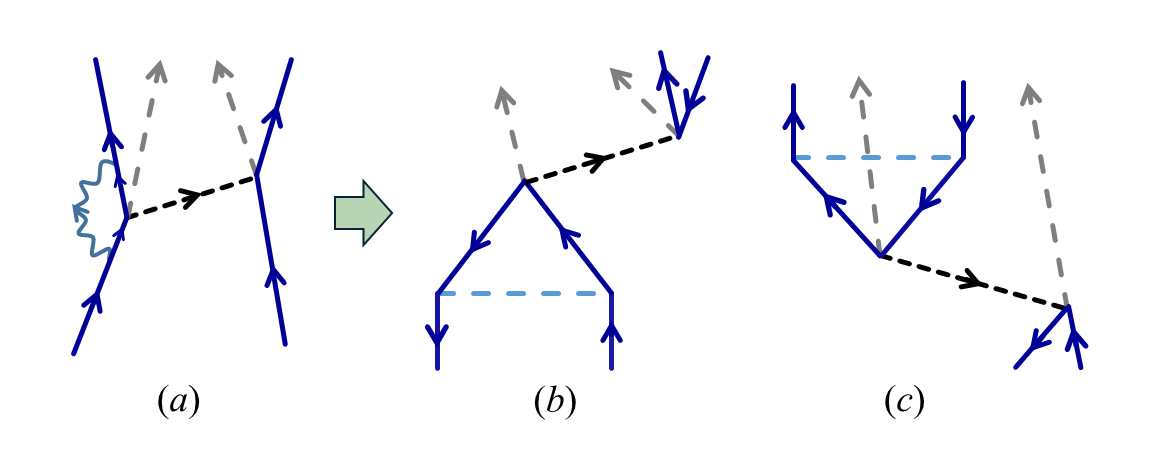}
\caption{ \protect \label{fig:0vbb_vc} \baselineskip=13pt 
Diagrams of the lowest-order vertex correction. The wavy line of (\textit{a}) stands for the meson, and the horizontal long-dashed line of (\textit{b}) and (\textit{c}) implies a nucleon-nucleon potential. The other parts are the same as those in Fig.~\ref{fig:0vbb_0th}. The wide arrow indicates a relation between (\textit{a}) and the other diagrams; see text.}
\end{figure}

The general equation of the decay rate of the $0\nu\beta\beta$ decay \cite{Pri59,Doi85} is given by
\begin{eqnarray}
D_{0\nu}=G_{0\nu} g_A^4 |M_{0\nu} |^2 \left(\frac{\langle m_\nu \rangle}{m_e}\right)^2,\langle m_\nu \rangle =\left|\sum_{i=1,2,3}U_{ei}^2 m_i \right|. \label{eq:decay_rate}
\end{eqnarray}
For the unperturbed calculation, the unperturbed NME
\begin{eqnarray}
M_{0\nu}^{(\textrm{unp})}=4\pi R\sum_B \sum_{pp^\prime nn^\prime} \langle pp^\prime\ |V_\nu (r_{12},E_B )| nn^\prime \rangle\langle F|c_{p^\prime}^\dagger c_{n^\prime} | B\rangle  \langle B |c_p^\dagger c_n | I \rangle,
\label{eq:M0v_unp}
\end{eqnarray}
is substituted for $M_{0\nu}$. $G_{0\nu}$ is the phase space factor originating from the emitted electrons \cite{Kot12}, and $m_e$ is the electron mass. $U_{ei}$ is the PMNS matrix element, and $m_i$ denotes the neutrino eigenmass. $R$ is the nuclear root-mean-square radius, $p$ and $n$ denote the proton and the neutron, respectively, and their creation and annihilation operators are expressed by $c^\dagger$ and $c$, respectively. $|I\rangle$, $|B\rangle$, and $|F\rangle$ are the initial, intermediate, and final nuclear states, respectively. $|I\rangle$ and $|F\rangle$ are the ground states of those nuclei. The two-body operator of the $0\nu\beta\beta$ decay can be approximated as
\begin{eqnarray}
&& V_\nu (r_{12},E_B )=h_+ (r_{12} )(-\bm{\sigma}_1\cdot\bm{\sigma}_2 + g_V^2/g_A^2 )  \tau_1^- \tau_2^-, \\[5pt]
&& h_+ (r_{12} )=\int \frac{d^3 \bm{q}}{(2\pi)^3} \frac{1}{|\bm{q}|} 
\frac{\mathrm{exp}[i\bm{q}\cdot(\bm{r}_1 - \bm{r}_2 )]}{(1/2) (M_I + M_F ) - \overline{E}_C - |\bm{q}| }.
\end{eqnarray}
The distance between two nucleons is denoted by $r_{12}$ $\equiv$ $|\bm{r}_1 - \bm{r}_2 |$. The Pauli spin matrix and the charge change operator of $n \rightarrow p$ are denoted by $\bm{\sigma}_1$ and $\tau_1^- $ for particle 1, respectively. The strength of the Fermi transition operator is $g_V$. The $\bm{\sigma}_1 \cdot \bm{\sigma}_2 \tau_1^- \tau_2^-$  $(\tau_1^- \tau_2^-)$ term causes the double-GT (double-Fermi) transitions. $M_I$ and $M_F$ are the nuclear masses of the initial and final nuclei, respectively, and $\overline{E}_C$ denotes the average of the intermediate-state energy $E_B$. The momentum of the neutrino is denoted by $\bm{q}$. The unpertubed process is represented by Fig.~\ref{fig:0vbb_0th}.

The lowest-order vertex correction (VC) is illustrated in Fig.~\ref{fig:0vbb_vc}(\textrm{a}), where the wavy line stands for the meson. The diagrams of Fig.~\ref{fig:0vbb_vc}(\textit{b}) and \ref{fig:0vbb_vc}(\textit{c}), denoted by \textit{JJV} and \textit{VJJ}, respectively, are obtained by replacing the meson by a nucleon-nucleon potential, and these are calculated. The equation of the \textit{VJJ} term is obtained by using the higher-order Rayleigh-Schr\"{o}dinger perturbation and a few approximations as
\begin{eqnarray}
M_{0\nu}^{(VJJ)}&=&4\pi R\sum_{j^\prime l^\prime mn} \langle j^\prime m |V_\nu (r_{12},\overline{E}_C )| l^\prime n\rangle \sum_{\substack{ ijkl \\ \textrm{ all different} } } \frac{1}{\varepsilon_i+\varepsilon_j+\varepsilon_k+\varepsilon_l } \frac{1}{2} \sum_{ijkl} \langle ij|:V:|lk\rangle \nonumber \\
& &V_{j-j}  V_{i-i}  U_{ll}  U_{kk} \left( U_{j^\prime -j}^\ast V_{l^\prime k}^\ast-U_{j^\prime k}^\ast V_{l^\prime-j}^\ast \right) \sum_{C_F C_I} \langle F|a_{-i} a_l | C_F\rangle U_{mm}^\ast V_{n-n}^\ast \langle C_F|C_I \rangle \nonumber \\ 
& &\langle C_I |a_m^\dagger a_{-n}^\dagger |I\rangle. \label{eq:M0v_VJJ}
\end{eqnarray}
Here, $V$ denotes the interaction for the perturbation corresponding to the horizontal long-dashed line in Fig.~\ref{fig:0vbb_vc}(\textit{b}) and (\textit{c}); the interaction to obtain the nuclear states is used. Some of the components do not contribute to the VC due to the structure of the equations. $V_{i-j}$, $U_{lk}$, and others are the transformation matrix elements between the single particles of the canonical basis \cite{Rin80} $(i,j,\cdots)$ and the canonical-quasiparticles (the indexes shared); $a_i^\dagger$ creates the canonical quasiparticle. Note that $U_{ij}=U_{ii}\delta_{ij}$ and $V_{i-j}=V_{i-i}\delta_{ij}$. The state with inverted angular momentum is denoted by $-i$. The diagonal matrix element of the canonical-quasiparticle energy matrix is denoted by $\varepsilon_i \equiv \varepsilon_{ii}$ and others. The normally ordered $:V:$ is defined with the canonical-quasiparticle basis. $|C_I\rangle$ and $|C_F\rangle$ are the intermediate states obtained by the quasiparticle random-phase approximation (QRPA) \cite{Rin80, Suh07} based on the initial and final ground states, respectively. For the VC terms, I use the Hartree-Fock-Bogoliubov (HFB) \cite{Rin80} ground states. The \textit{JJV} term $M_{0\nu}^{(JJV)}$ can be obtained analogously. The VC term is the sum of $M_{0\nu}^{(VJJ)}$ and $M_{0\nu}^{(JJV)}$, and for the perturbed calculation, the sum of $M_{0\nu}^{(\textrm{unp})}$ [Eq.~(\ref{eq:M0v_unp})] and the VC term is inserted to $M_{0\nu}$ in the equation of $D_{0\nu}$ [Eq.~(\ref{eq:decay_rate})]. The equations for the two-neutriono double-$\beta$ ($2\nu\beta\beta$) NME are the same as those for the $0\nu\beta\beta$ NME, except that the neutrino potential $h_+ (r_{12})$ is replaced by the inverse of the energy denominator without $|\bm{q}|$ and the average of $E_B$ is reduced to $E_B$.

\section{\label{sec:results} Result and Discussion}
The nuclear states were obtained by the QRPA with the Skyrm interaction (SkM$^\ast$) \cite{Bar82} and the contact isovector $pp$, $nn$, $pn$, and the isoscalar $pn$ pairing interactions. For technical details of the QRPA and HFB calculations, see Ref.~\cite{Ter19}. The obtained NME components for $^{136}$Xe $\rightarrow$ $^{136}$Ba are shown in Table \ref{tab:0th_vc}. The intermediate nucleus is $^{136}$Cs. It is seen that the VC reduces the NME by nearly 30\%, and the \textit{VJJ} term has the main contribution. The change in the Fermi component is significant. According to the analysis of the details, two-particle two-hole components of the final state mainly cause the VC. 

Usually, the effective $g_A$ is introduced to reproduce an experimental half-life with the unperturbed transition operator. It is possible to simulate this definition using the NME with the perturbed transition operator and the bare $g_A$ as a substitute for the experimental value. I use $g_V$ = 1 according to the usual method. The effective $g_A$ obtained in this manner [$g_A^\textrm{eff}(\textrm{pert})$] is summarized in Table \ref{tab:gAeff_pert} for the $0\nu\beta\beta$ and $2\nu\beta\beta$ NMEs. It is seen that the difference between the two $g_A^\textrm{eff}(\textrm{pert})$ is 10\%. This is the most striking finding of this study because it was thought that the effective $g_A$ for the $0\nu\beta\beta$ NME is unknown at all; this includes a possibility that the two effective $g_A$’s are significantly different. Now, my result indicates the possibility that the phenomenological effective $g_A$ to reproduce the experimental half-life of the $2\nu\beta\beta$ decay \cite{Bar19} can be approximately used for the calculation of the $0\nu\beta\beta$ NME. 
\begin{table}[t]
\caption{\label{tab:0th_vc} Unperturbed term [Eq.~(\ref{eq:M0v_unp})] and vertex correction (VC) for the GT and Fermi components of the $0\nu\beta\beta$ NME for $^{136}$Xe by the HFB and QRPA calculations. VC has two components, $JJV$ and $VJJ$, of which diagrams are shown in Fig.~\ref{fig:0vbb_vc} (\textit{b}) and (\textit{c}), respectively. See also Eq.~(\ref{eq:M0v_VJJ}). The NME is calculated according to
(GT) $-$ ($g_V$/$g_A$)$^2$ (Fermi) with $g_A$ = 1.27 and $g_V$ = 1.0.}
\begin{ruledtabular}
\begin{tabular}{ccrrr}
\multicolumn{2}{c}{Term} & GT & Fermi & NME \\
\hline \\[-11pt]
\multicolumn{2}{l}{Unperturbed} & 3.095 & $-0.467$ & 3.384 \\[2pt]
\multirow{2}{*}{VC} & $JJV$ & $-0.048$  & 0.025 & $-0.064$ \\[2pt]
 & $VJJ$ & $-0.715$ & 0.519 & $-1.037$ \\[2pt]
\multicolumn{2}{l}{$JJV+VJJ$} & $-0.763$ & 0.544 & $-1.100$ \\[2pt]
\multicolumn{2}{l}{Unperturbed + VC} & 2.332 & 0.077 & 2.284 \\[2pt]
\multicolumn{2}{l}{VC/unperturbed (\%)} & $-24.6$ & $-116.5$ & $-32.5$ \\[2pt]
\end{tabular}
\end{ruledtabular}
\end{table}
%
\begin{table}
\caption{\label{tab:gAeff_pert} Effective $g_A$ obtained referring to the NME with the perturbed transition operator. Those for the $0\nu\beta\beta$ and $2\nu\beta\beta$ NMEs for $^{136}$Xe are shown.}
\begin{xtabular*}{0.3\textwidth}{c@{\extracolsep{\fill}}r}
\hline\\[-14pt]
\hline\\[-13pt]
  & $g_A^\textrm{eff}(\textrm{pert})$ \\[3pt]
\hline \\[-12pt]
$0\nu\beta\beta$ & 1.019 \\[2pt]
$2\nu\beta\beta$ & 0.891 \\[2pt]
\hline\\[-14pt]
\hline
\end{xtabular*}
\end{table}


\begin{acknowledgments}  
This study was supported by the Czech Science Foundation (GA\v{C}R), project No.~24-10180S. The computation for this study was performed by Karolina, IT4Innovations supported by the Ministry of Education, Youth and Sports of the Czech Republic through the e-INFRA CZ (ID:90254); the computers of MetaCentrum provided by the e-INFRA CZ project (ID:90254) and supported by the Ministry of Education, Youth and Sports of the Czech Republic; Yukawa-21 at Yukawa Institute for Theoretical Physics, Kyoto University. 
\end{acknowledgments}
%
\vfill
\end{document}